\documentclass[conference,a4paper]{IEEEtran}
\IEEEoverridecommandlockouts
\usepackage{cite}
\usepackage{amsmath,amssymb,amsfonts}
\usepackage{algorithmic}
\usepackage{graphicx}
\usepackage{textcomp}
\usepackage{xcolor}
\usepackage{slashbox}
\usepackage{tikz}
\usepackage{booktabs}
\usepackage{caption}
\usepackage[a4paper,top=1.9cm,bottom=4.29cm,left=1.29cm,right=1.29cm]{geometry}
\def\BibTeX{{\rm B\kern-.05em{\sc i\kern-.025em b}\kern-.08em
    T\kern-.1667em\lower.7ex\hbox{E}\kern-.125emX}}

\graphicspath{{./pictures/}}
\usepackage{blindtext,graphicx}
\usepackage[absolute]{textpos}
\begin{document}
\begin{textblock}{17}(2,0.5)
	\noindent \centering {\small This work has been submitted on October 28 2020 to the IEEE ICC 2021 conference for possible presentation and
subsequent publication by the IEEE. Copyright may be transferred without notice, after which this version may no
longer be accessible.}
\end{textblock}

\title{Extending the Lora modulation to add further parallel channels and improve the LoRaWAN network performance  \\
\thanks{Lorenzo Vangelista acknowledges the economical support by the POR FESR 2014-2020 Work Program of the Veneto Region (Action 1.1.4) through the project No.10066183 titled “Sistema domotico IoT integrato ad elevata sicurezza informatica per smart building''. Alessandro Cattapan was a student at University of Padova when working on this paper}
}

\author{\IEEEauthorblockN{Lorenzo Vangelista}
\IEEEauthorblockA{\textit{Department of Information Engineering} \\
\textit{University of Padova}\\
Padova, Italy \\
lorenzo.vangelista@unipd.it}
\and
\IEEEauthorblockN{Alessandro Cattapan}
\IEEEauthorblockA{\textit{Department of Information Engineering} \\
\textit{University of Padova}\\
Padova, Italy \\
alessandro.cattapan.2@studenti.unipd.it}
}

\maketitle

\begin{abstract}
In this paper we present a new modulation, called DLoRa, similar in principle to the conventional LoRa modulation and compatible with it in terms of bandwidth and numerology. DLoRa departs from the conventional LoRa modulation as it is using a decreasing instantaneous frequency in the chirps instead of an increasing one as for the conventional LoRa modulation.
Furthermore we describe a software environment to accurately evaluate  the ``isolation'' of the different virtual channels created both by LoRa and DLoRa  when using different Spreading Factors. Our results are in agreement with the ones present in literature for the conventional LoRa modulation and show that it is possible to double the number of channels by using simultaneously LoRa and DLora. The higher (double) number of subchannels available is the key to improve the network level performance of LoRa based networks. 
\end{abstract}

\begin{IEEEkeywords}
LoRa Modulation, LoRaWAN system, Digital
Modulation, Internet of Things
\end{IEEEkeywords}

\section{Introduction}
The Internet of Things (IoT) connectivity has seen a significant change of paradigm in recent years.   The usual paradigm some years ago was that of the ``Wireless Sensor Networks'' (WSN,) in which the mesh topology was almost always assumed.  Nowadays the ``Low Power Wide Area Networks'' is the most widespread paradigm \cite{centenaro2016, raza2017}. This is true both for the licensed frequency bands, where the NB-IoT is the technology of choice, and the unlicensed frequency bands where many different technologies are used, such as SigFox, LoRaWAN etc. 

The most popular technology for the LPWAN networks in unlicensed frequency bands is LoRaWAN, based on the LoRa modulation, a proprietary technology owned by Semtech, and standardized from the protocol side by the LoRa Alliance \cite{centenaro2016}. 

In this paper we focus on the LoRaWAN technology and especially on the underlying LoRa modulation. One its main challenges is, of course, to maximize the throughput. The LoRa modulation is an adaptive modulation for which one can select a peculiar parameter, known as Spreading Factor (SF), and by using different SFs one can create almost parallel, quasi orthogonal, channels in the same band. The amount of these almost orthogonal channels is limited and corresponds to setting the SF in the range $7,8, \ldots 12$, for a total of 6 almost orthogonal channels. To increase the throughput, increasing this number would be of course beneficial but the current definition of the LoRa modulation cannot accommodate such an increase.         

In this paper, as a first novel contribution, we  define  a new variation of the LoRa modulation, called DLoRa, which enable to create a new set of 6 almost parallel channels, indexed once again by the SF in the range $7,8, \ldots 12$,  which are not only quasi orthogonal to each other but also quasi orthogonal to the ones using the original LoRa modulation. In total then, by using the original LoRa modulation and DLoRa one can have a set of 12 quasi orthogonal channels, with an increase of a factor of 2 with respect to the original LoRa modulation. 

The availability of a higher number of quasi orthogonal channels is, of course, beneficial for the network performance; as a matter of fact, one can think of splitting the End Nodes (EN) which in the original LoRaWAN network based on LoRa are using a certain SF in two parallel channels. This split can be done for every SF, greatly contributing to enhance the throughput since LoRaWAN uses basically (for class A devices\footnote{For the definition of Class A devices /End Nodes see \cite{centenaro2016}.}, which are the vast majority) a simple Aloha access protocol.  

A second novel contribution of this paper is related to the quantitative   evaluation of the aforementioned quasi orthogonality. As matter of fact, this in many papers (e.g., \cite{goursaud2015}) has been done either in a theoretical way or by numerical simulation which the paper \cite{tinnirello} has been proved to inaccurate. The authors in \cite{tinnirello} resorted to a better simulation environment then and eventually to the use of hardware--based experimental evaluation. In this paper we propose and describe an accurate simulation model for the quantitative   evaluation of the  quasi orthogonality between different SFs (including the ones related to the newly proposed modulation), which results are very similar to the simulator results of Table I of \cite{tinnirello}. The evaluation confirms that the channels with conventional LoRa modulation and with DLoRa are all quasi-orthogonal. 

The paper is organized as follows: In Section \ref{sec:notation} the system model and the notation is introduced for the conventional LoRa modulation. In Section \ref{sec:DLoRa} the newly introduced DLoRa modulation and its properties are described. I Section \ref{sec:isolation} our approach to quantify the quasi--orthogonality (also called from now on ``isolation'') between the channels with different SF and different modulation (conventional LoRa and DLoRa) is described. The obtained results are then presented. In Section \ref{sec:conclusions} the conclusions of our work are drawn

\section{System model, notation and related work}
\label{sec:notation}
In this paper we follow the notation of \cite{chiani}. For the plain LoRa M-ary modulation ($M=2^{SF}$ where $SF$ is the so--called Spreading Factor), in the band $[f_0-B/2,f_0+B/2]$ the complex envelope of transmitted signal in the interval $[0,T_s)$ ($T_s=MT_c$ with $T_c=1/B$) 
is, for a transmitted symbol $a\in \{0, 1, \ldots M-1\}$:
\begin{IEEEeqnarray}{lCr}
	x(t;a)&\!=\!&\exp\!\left\lbrace \! \jmath 2 \pi Bt \! \left(\! \frac{a}{M} \! - \! \frac{1}{2} \! + \! \frac{Bt}{2M} \!\right) \! \!- \! u \! \left(\! t \!- \!\frac{M-a}{B}\right) \!\right\rbrace  \label{eq:symbol_def_plain}
\end{IEEEeqnarray}

\noindent where$u(t)$ is the unit step function.

Referring to \textit{Property 3} of \cite{chiani}, the waveforms \eqref{eq:symbol_def_plain} for different values of $a$ are orthogonal in the discrete time domain and orthogonal with a very good approximation in the continuous time domain (see \cite{chiani} for the analytical expression in the continuous time domain). 

The orthogonality conditions were analyzed in \cite{chiani} assuming the same SF and they form the basis of the LoRa modulation. However, using the LoRa modulation in LoRaWAN networks entails the usage of several different SFs by several different ENs. This usage leads to the need of having (ideally) orthogonality between packets (i.e. groups of subsequent symbols) 
\begin{itemize}
	\item with different spreading factors;
	\item  not synchronized each other.
\end{itemize}
Unfortunately this (ideal) orthogonality, which is often assumed, does not hold and has been pointed out since the publication of  one of the first paper on LoRaWAN i.e., \cite{goursaud2015}.
Quoting \cite{tinnirello}
\begin{quote}
	Current studies [...] assume
that the utilization of multiple transmission channels and
SFs lead to a system that can be considered as the simple
super-position of independent (single channel, single SF) subsystems.
This is actually a strong simplification, especially
because the SFs adopted by LoRa are quasi-orthogonal
\end{quote}

Taking into account the above mentioned quasi orthogonality, algorithms at network management level to assign the different SFs to the ENs have been proposed by many authors (see e.g., \cite{lim2018}), the latest and most promising one being \cite{bianchi}. In the vast majority of them the availability of a higher number of channels (even if quasi orthogonal) is definitely beneficial. This is remarking how valuable is to double the number of channels.

\section{The DLoRa modulation}
\label{sec:DLoRa}

The conventional LoRa modulation is using a Chirp--like modulation where the instantaneous frequency of the chirps is always increasing. The key idea novel ideal of the DLora modulation is extending to the LoRa modulation using a decreasing instantaneous frequency. The expression \ref{eq:symbol_def_plain} becomes then for a transmitted symbol $a\in \{0, 1, \ldots M-1\}$:
\begin{IEEEeqnarray}{lCr}
	\!\!\!\!\!x_D(t;a)&\!=\!&\exp\!\left\lbrace \! \jmath 2 \pi Bt \! \left(\! \frac{a}{M} \! - \! \frac{1}{2} \! - \! \frac{Bt}{2M} \!\right) \! \!+ \! u \! \left(\! t \!- \!\frac{M-a}{B}\right) \!\right\rbrace  \label{eq:dlora}
\end{IEEEeqnarray}

\begin{figure*}[h]
	\centering
	\includegraphics[width=0.5\linewidth]{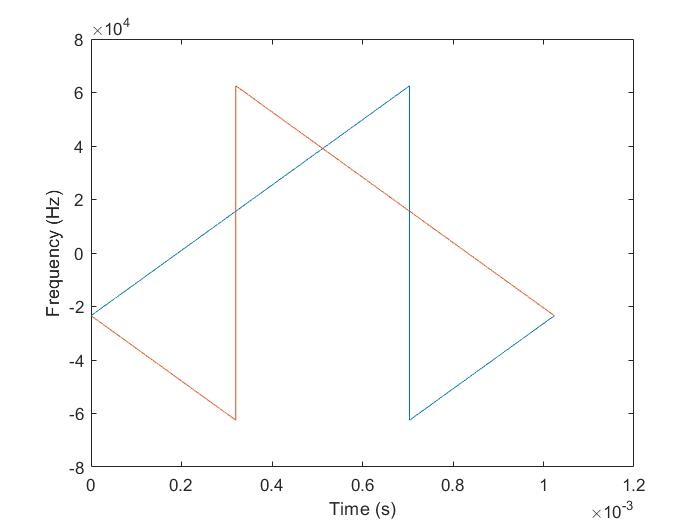}
	\caption{Frequency behaviour of for LoRa (blue) and DLoRa (orange) signal with SF = 7 and a = 40.}
	\label{fig:dilorafrequency}
\end{figure*}

In Fig\ref{fig:dilorafrequency} a plot is shown for the DLoRa modulation compared to the LoRa conventional modulation.

It is straightforward to show that the DLoRa modulation enjoys the same properties, as far as the orthogonality of the waveforms in equation (\ref{eq:dlora}) is concerned, as for the conventional LoRa modulation. It is as well straightforward to see that the demodulation of the DLoRa signals can be obtained using the demodulator structure of the one for the LoRa signals, just replacing a downchirp instead of an upchirp.

\begin{table*}[]
		\centering
	\begin{tabular}{|c|c|c|c|c|c|c|c|c|c|c|c|c|}
		\hline
		\backslashbox{SF ref}{SF int}   & $7$ & $8$ & $9$ & $10$ & $11$ & $12$ & $7_D$ & $8_D$ & $9_D$ & $10_D$ & $11_D$ & $12_D$ \\ \hline
		$7$  & 0       & -10     & -12     & -12      & -13      & -14      & -11     & -11     & -11     & -11      & -12      & -13      \\ \hline
		$8$  & -12     & 0       & -13     & -14      & -15      & -16      & -13     & -14     & -14     & -14      & -14      & -15      \\ \hline
		$9$  & -16     & -15     & 0       & -16      & -17      & -18      & -15     & -16     & -16     & -17      & -17      & -17      \\ \hline
		$10$ & -18     & -18     & -18     & 0        & -19      & -20      & -18     & -18     & -19     & -19      & -20      & -20      \\ \hline
		$11$ & -21     & -21     & -21     & -21      & 0        & -21      & -21     & -21     & -21     & -22      & -22      & -23      \\ \hline
		$12$ & -23     & -24     & -24     & -24      & -24      & 0        & -23     & -24     & -24     & -24      & -25      & -25      \\ \hline
		$7_D$  & -11     & -11     & -11     & -11      & -12      & -13      & 0       & -10     & -12     & -12      & -13      & -14      \\ \hline
		$8_D$  & -13     & -14     & -14     & -14      & -14      & -15      & -12     & 0       & -13     & -14      & -15      & -16      \\ \hline
		$9_D$  & -16     & -16     & -17     & -17      & -17      & -17      & -16     & -15     & 0       & -16      & -17      & -18      \\ \hline
		$10_D$ & -18     & -18     & -19     & -20      & -20      & -20      & -18     & -18     & -18     & 0        & -19      & -20      \\ \hline
		$11_D$ & -21     & -21     & -21     & -22      & -22      & -23      & -21     & -21     & -21     & -21      & 0        & -21      \\ \hline
		$12_D$ & -23     & -24     & -24     & -24      & -25      & -25      & -23     & -24     & -24     & -24      & -24      & 0        \\ \hline
	\end{tabular}
	\caption{Thresholds on the SIR between the desired signal and the interferer signal in order to guarantee a $BER = 0.01$. The subscript $D$ indicates the DLoRa modulation}
	\label{tab:SIR_thresholds}
\end{table*}

\section{Isolation between Spreading Factors}
\label{sec:isolation}

 In \cite{goursaud2015} we the isolation between the packets transmitted at different SFs is provided in terms of Signal to Interfence plus Noise Ration (SINR).
The main problem with the values provided in \cite{goursaud2015} is that they are provided without any explanation on how they are obtained.
Moreover these values are quite different from the one  values in therms of Signal to Interference (SIR) obtained in  \cite{tinnirello}, where the results are first obtained through MATLAB simulations and then partially confirmed by USRP (Universal Software Radio Peripheral) experiments.
Even though the Matlab's values in \cite{tinnirello} are not perfectly matching the one obtained in practice, the two results can be considered a good approximation of the real case.

In this Section we will provide a description and the results of the procedure we have used to estimate the values of the SIR between two packets with different SF and possibly different modulation (conventional LoRa and DLoRa)  via a Matlab program, inspired by the work of \cite{tinnirello}.

\subsection{Setting of the Simulations in Matlab}

For simplicity from now on we will consider only two packets, since that the results can be easily generalized to the more common case of multiple packets. 
One packet will be considered as the \textit{reference packet}, which is the packet that we desire to survive the interference, while the other one will be called \textit{interfering packet}.

 The two packets interferes with each other when they are overlapping in time. This overlap can be  partial or  complete.
Obviously the results from these two situations are different.
However, in our simulations, in order to have a common base to analyze and compare the results, we decided to take into consideration only the cases where the interfering packet is  completely overlapping with the reference packet.

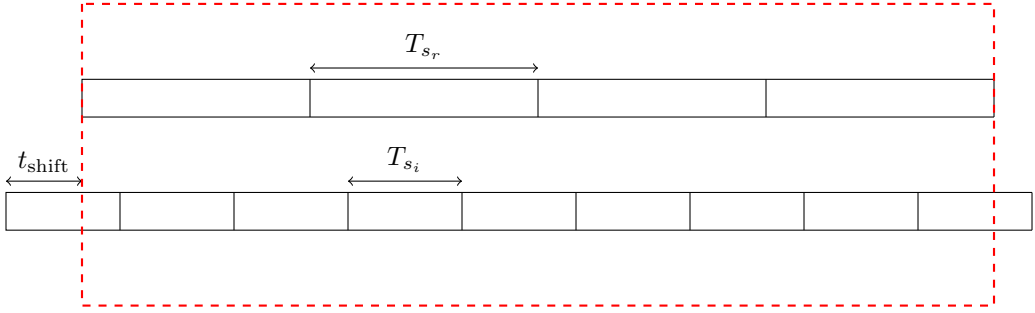
\begin{figure*}
	\centering
\scalebox{1}{
	
	\begin{tikzpicture}
	
	\draw[<-] (4,0.15) -- (5.5,0.15) node[anchor=south] {$T_{s_r}$};
	\draw[->] (5.5,0.15) -- (7,0.15);
	
	\draw[<-] (4.5,-1.35) -- (5.25,-1.35) node[anchor=south] {$T_{s_i}$};
	\draw[->] (5.25,-1.35) -- (6,-1.35);
	
	\draw[<-] (0,-1.35) -- (0.5,-1.35) node[anchor=south] {$t_{\mathrm{shift}}$};
	\draw[->] (0.5,-1.35) -- (1,-1.35);
	
	\draw (13,0) -- (1, 0) -- (1,-0.5) -- (13,-0.5);
	\foreach \x in {4,7,10,13}
	\draw (\x,0) -- (\x,-0.5);
	
	\draw (13.5,-1.5) -- (0, -1.5) -- (0,-2) -- (13.5,-2);
	\foreach \x in {1.5,3,4.5,6,7.5,9,10.5,12,13.5}
	\draw (\x,-1.5) -- (\x,-2);
	
	\draw[red,thick,dashed] (1, 1) rectangle (13,-3);
	
	\end{tikzpicture}
	
}
	\caption{Two completely colliding packets and the analysis window (in red).}
	\label{fig:InterferingPacketVisualization}
\end{figure*}

The target situation for our simulations is shown in Figure \ref{fig:InterferingPacketVisualization}, where the red rectangle represent the region in which we are going to analyze the interference between the two packets.
The packet above is the desired packet while the one below is the colliding packet. 
From the figure it is clear that the two packets have different spreading factors in fact the chirp duration of the first one $T_{s_r}$ is twice the duration of the chirps of the second packet $T_{s_i}$.

Given the length (in bytes) of the reference packet's payload as a parameter $N_{bytes}$, the number of bits $N_{bits_r} $ in the packet is computed as
\begin{equation}
N_{bits_r} = \left\lceil \frac{8N_{bytes}}{(4 N_{bs_r})}\right\rceil \cdot (4 N_{bs_r})
\label{eq:NBitsReferencePacket}
\end{equation}
where $N_{bs_r}$ corresponds to the number of bits contained in each symbol, which in our case corresponds to Spreading Factor of the reference packet, $SF_r$. From (\ref{eq:NBitsReferencePacket}) it is easy to compute the number of chirps $ N_{chirp_r}  $ inside every packet
\begin{equation}
N_{chirp_r} = \left\lceil \frac{N_{bits}(CR+4)}{4N_{bs_r}} \right\rceil
\label{eq:NChirpsReferencePacket}
\end{equation}
where $4/(CR+4)$ is the coding rate with $CR \in \{0,1,2,3\}$. 

The time on air $ ToA $ of the packet is
\begin{equation}
ToA = N_{chirp_r} \cdot M_r
\end{equation}
The total number of chirp that the interfering packet should have to have a complete overlap between the two packets is then 
\begin{equation}
N_{chirp_i} = \left\lceil \frac{ToA}{M_i} \right\rceil + 1
\end{equation}
where the $+1$ is added to compensate the effects due to the shift of the interfering packet with respect to the desired one.

\begin{figure*}
	\centering
\scalebox{1}{
	
	\begin{tikzpicture}
	
	\draw[<-] (4.5,1.2) -- (5.25,1.2) node[anchor=south] {$T$};
	\draw[->] (5.25,1.2) -- (6,1.2);
	
	\draw (13,0) -- (3, 0) -- (3,-1) -- (13,-1);
	\foreach \x in {4.5,6,7.5,9,10.5,12}
	\draw (\x,0) -- (\x,-1);
	
	\draw (12,0) -- (13,0);
	\draw (12,-1) -- (13,-1);
	
	\draw (13,-2) -- (0.3, -2) -- (0.3,-3) -- (13,-3);
	\foreach \x in {1.5,3,4.5,6,7.5,9,10.5,12}
	\draw (\x+0.3,-2) -- (\x+0.3,-3);
	
	\draw[red,thick,dashed] (13,1) -- (3, 1) -- (3,-4) -- (13,-4);
	
	\foreach \x in {4.5,6,7.5,9,10.5,12}
	\draw[red,thick,dashed] (\x,1) -- (\x,-4);
	
	\draw[<-] (0.3,-1.8) -- (1.05,-1.8) node[anchor=south] {$t_{int}$};
	\draw[->] (1.05,-1.8) -- (1.8,-1.8);
	
	\draw[<-] (1.8,-1.8) -- (2.4,-1.8) node[anchor=south] {$t_{float}$};
	\draw[->] (2.4,-1.8) -- (3,-1.8);
	
	\end{tikzpicture}
	
}
	\caption{Alignment of two interfering packets. The packet above is the reference packet, which is synchronized with the receiver (represented in red). While the packet below is the interfering one and it is randomly shifted.}
	\label{fig:InterferingPacketsShiftRapresentation}
\end{figure*}
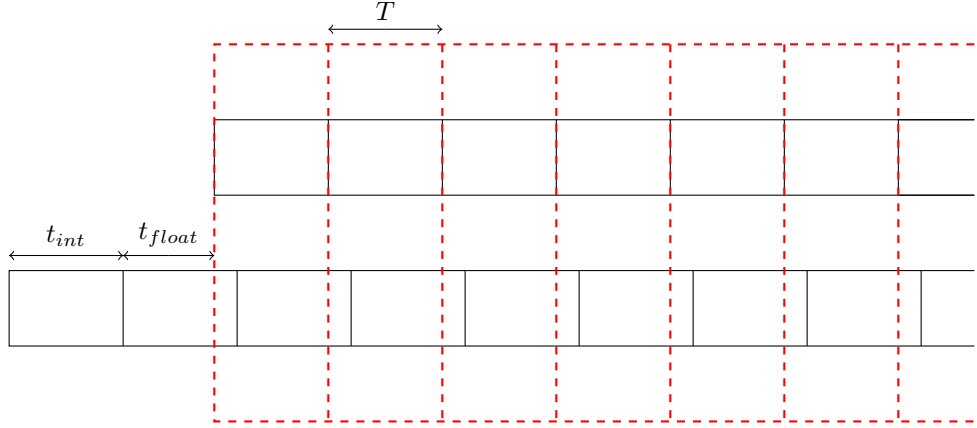

To explain how the interfering packet is shifted we need to consider Figure \ref{fig:InterferingPacketsShiftRapresentation} that represent the reference packet, the one above, and the interfering packet, the one below.
The sampling time (in multiples of $T=1/B$ seconds) of the receiver is drawn in red and, as we can see from the figure, it is perfectly synchronized with the sampling time of the reference packet.
Clearly the receiver is sampling the signal asynchronously with respect to  the interfering packet.
In particular the interfering packet is shifted randomly by
\begin{equation}
t_{shift} = t_{int} + t_{float}
\end{equation}
where $t_{int}$ is a shift of an integer number of samples, that is $t_{int}=n_{int}T$ with $n_{int} \in \{0,1,2,...,M-1\}$ and this is the real reason why we have previously added one to the $N_{chirp_i}$.
The  shift $f_{float}$ is instead a fractional part of one chip period and is defined as 
\begin{equation}
t_{float} = n_{float}\frac{B}{SR}, \quad\quad n \in \{0,1,2,...,SR-1\}
\end{equation}
where $SR$ is a variable parameter.
Every time that we simulate two new interfering packets we randomly generate $t_{shift}$ by randomly drawing $n_{int}$ and $n_{float}$ from their respective set of values. 
The setting of the SIR is done by tuning the signals amplitude $A_r$ and $A_i$
\begin{equation}
SIR = 10 \log_{10} \frac{P_r}{P_i} = 10 \log_{10} \frac{A^2_r}{A^2_i} 
\end{equation}
More in detail, the two packets are built in such a way to have unitary power, then the reference packet is multiplied by $A_r=1$, while the interfering packet is multiplied by $A_i$ computed as follow
\begin{equation}
A_i = \frac{A_r}{10^{(\frac{SIR}{20})}}
\end{equation}
To speed up the simulations the computation of the interfering signal shifted by $t_{float}$ is done using the continuous time equation of the LoRa chirps in (\ref{eq:symbol_def_plain}) for the conventional LoRa modulation and (\ref{eq:dlora}) for the DLora Modulation sampled only on the time instants of our interest, which means the sampling instants of the receiver (the vertical dashed red lines in Figure \ref{fig:InterferingPacketsShiftRapresentation})
\begin{table}[]
	\hspace*{1cm}
	\centering
	\begin{tabular}{@{}ll@{}}
		\toprule
		\textbf{Parameter}                                        & \textbf{Value}             \\ \midrule
		Bandwidth (kHz)                                           & 125                        \\ \midrule
		Coding Rate (CR)                                          & 4/5                        \\ \midrule
		Message Size (bytes)                                      & 20                         \\ \midrule
		SIR range                                                 & [-30 dB, 10 dB]            \\ \midrule
		SIR step                                                  & 1 dB                       \\ \midrule
		$BER_{target}$                                            & 0.01                       \\ \midrule
		SR                                                        & 100                        \\ \midrule
		Minimum number of errors                                                         &                       \\
		to be observed in each SIR level & 100                        \\ \bottomrule
	\end{tabular}
	\caption{Simulation Parameters.}
	\label{tab:SimulationsParametersValue}
\end{table}

\subsection{Results of the simulations}

We have run then simulations for every possible spreading factors combination (including both conventional LoRa modulation and DLoRa modulation) iterating  over the SIR  in the range$[-30, 10]dB$ with a step of $1dB$.
For each SIR level we simulated multiple collisions of the two packets, whose content is randomly generated. 
The interfering packet is randomly shifted every time with a different $t_{shift}$.
After every block of simulations  the corresponding BER and PER at the specific SIR is computed, iterating  until we reach the target $BER_{target}=0.01$.
The simulations parameters are summarize in Table \ref{tab:SimulationsParametersValue}.

 The results of the simulations are reported in Table \ref{tab:SIR_thresholds}.  We can see from Table \ref{tab:SIR_thresholds} that:
 \begin{itemize}
 	\item the results of the simulations are in agreement (within a 1-2 dB) with simulations results of \cite{tinnirello};
 	\item the DLoRa modulation exhibits a good isolation (at least -11dB) from the conventional LoRa modulation: this provides a new set of quasi orthogonal channels with respect to the conventional LoRa modulation; this new set of channels is to be used to diminish the load on the conventional channels and improve the throughput.
 \end{itemize}
 
\section{Conclusions}
\label{sec:conclusions}

In this paper we have introduced a new variation of the LoRa modulation, called DLoRa, which uses a decreasing instantaneous frequency (instead of an increasing one as for the conventional LoRa modulation). We have then described a simulation environment, built with Matlab, which allows a detailed assessment of the isolation of the different sub--channels (i.e., channels using different SFs). We tested it and the results are in agreement with results provided in literature. We have then tested with the aforementioned software environment the isolation among the DLoRa channels and the conventional LoRa channels for all the different combination of SFs. The results demonstrate that it is possible to use at the same time LoRa and DLora modulations, thus doubling the number of channels from 6 to 12. This is enabling a superior network level performance since the End Nodes can be assigned to twice the number of subchannels, i.e., we can have half the number of End Nodes per subchannel. The net result is that a reduction of the collision probability is possible. Further investigation are in progress to quantify the network level improvement of the introduction of DLoRa via our NS3 (see https://www.nsnam.org/ for more information on the NS3 network simulation tool) based network simulation environment \cite{NS3}.

\end{document}